\documentclass[12pt]{article}

\usepackage[a4paper,margin=25mm]{geometry}
\usepackage{newtxtext,newtxmath}
\usepackage{amsmath}
\usepackage{setspace}
\usepackage{booktabs}
\usepackage{tabularx}
\usepackage{array}
\usepackage{graphicx}
\usepackage{fancyhdr}
\usepackage[hidelinks]{hyperref}
\usepackage[style=apa,backend=biber,natbib=true,sorting=nyt]{biblatex}
\DeclareLanguageMapping{american}{american-apa}

\newcolumntype{Y}{>{\raggedright\arraybackslash}X}
\begin{document}

\begin{center}
{\Large\bfseries Measuring Human Resilience to Unreliable AI Assistance: A Stress-Test Framework for Human--AI Interaction}
\end{center}

\vspace{1em}
\noindent\textbf{Shaoshan Liu}\\
Shenzhen Institute of Artificial Intelligence and Robotics for Society (AIRS), Shenzhen, China\\
\textit{Correspondence: shaoshanliu@cuhk.edu.cn}

\vspace{0.8em}
\noindent\textbf{Running head:} HUMAN RESILIENCE TO UNRELIABLE AI

\vspace{1.0em}
\noindent\textbf{Abstract}

As AI increasingly mediates human decisions, users must remain effective not only when assistance works as expected but also when its quality deteriorates. Human--AI research has developed rich measures of trust, reliance, and calibration, while human--autonomy research has examined resilience at the team and system levels. A missing measurement level is the human decision maker: how much task function is preserved or recovered when AI advice becomes unreliable? We introduce the Resilience Stress Test, a challenge-conditioned HCI framework that measures performance relative to a prechallenge baseline and summarizes functional loss and recovery. Four construct-validation tests draw on five public human--AI studies comprising 1,074 analytic participants. A controlled decline in AI quality produced sharp degradation followed by partial adaptation; misleading advice reduced performance in art and dermatology; baseline adjustment substantially attenuated the apparent complexity gradient in trip planning; and a logical-reasoning tutorial improved self-assessment calibration while challenged performance remained stable. The framework makes human resilience to unreliable AI operational as a functional outcome distinct from competence, calibration, and reliance, and provides a prospective protocol for comparing interfaces under reproducible AI reliability stress.

\vspace{0.5em}
\noindent\textbf{Keywords:} human--AI decision making; unreliable AI; adaptive performance; appropriate reliance; resilience measurement

\newpage

\section{Introduction}

AI is becoming an intermediary in human judgment. Decision-support systems now recommend diagnoses, rank options, generate analyses, and propose actions that people may incorporate into consequential decisions. This creates a new human-performance problem. The relevant question is no longer only whether people can perform a task or whether AI can improve their average accuracy. People must also remain effective when the quality of the AI assistance changes. A system that is useful during normal operation can become a source of error when its recommendations deteriorate, while a reliance strategy learned under high reliability may become maladaptive after that change. Human-centered AI research has therefore emphasized reliable systems, meaningful human control, and evaluation under changing system capability \parencite{shneiderman2020hcai,xu2022transitioning,simkute2024ironies}.

This problem motivates resilience as a human--AI interaction construct. Recent work on human resilience in the AI era argues that as AI takes over tasks and mediates decisions, an increasingly important human capability is the ability to preserve agency and function as technological conditions change \parencite{liu2025human}. That broader argument leaves an HCI question unresolved: if resilience matters, how should it be observed and measured during actual interaction with AI? A useful measure must distinguish a person who was weak at the task before AI failure from one whose performance specifically collapses because the AI became unreliable. It must also distinguish resistance to bad advice from blanket rejection of AI, because refusing correct advice can destroy the benefit of assistance.

Existing HCI constructs capture important parts of this problem without directly measuring the functional response. Trust concerns a user's attitude toward the system, reliance describes whether AI advice enters the decision, and calibration links those responses to system capability \parencite{lee2004trust,schemmer2023reliance}. Automation-bias and appropriate-reliance research explains why incorrect recommendations are accepted or rejected \parencite{goddard2012automationbias,bucinca2021trust}. Recent IJHCI studies further show that transparency, calibration feedback, and assistance-delivery mechanisms can change how users engage with automated advice \parencite{tatasciore2025calibrating,nair2025solicit}. These measures explain the interaction process. They do not directly answer an outcome question that becomes central under failure: how much of the human task function survives, how deeply does it deteriorate, and does it recover as the user adapts?

Resilience is not new to human factors or human--machine systems. \textcite{hoffman2017measuring} argued for quantitative resilience measurement in complex human--machine work. Human--autonomy studies have examined how teams reorganize and recover after automation and autonomy failures \parencite{demir2019resilience,grimm2023dynamical}. The National Academies has likewise identified resilient human--AI performance under off-normal conditions and the development of practical resilience metrics as open research needs \parencite{nasem2022humanai}. This literature establishes the importance of resilience under technological failure, but its principal unit is the human--autonomy team or sociotechnical system. The human decision maker receiving AI advice creates a different measurement opportunity: the same task can be observed before AI advice, the correctness of that advice can be identified, and the human decision can be observed again afterward.

The present paper focuses on this missing decision level. We define \emph{demonstrated resilience} as preservation or restoration of a protected human function under a specified AI reliability challenge. To our knowledge, prior work has not operationalized this challenge-conditioned human outcome by combining (i) a prechallenge or pre-advice measure of human function, (ii) an identifiable change in AI assistance quality, and (iii) post-challenge human performance, while empirically separating the resulting measure from baseline competence, reliance, and calibration across AI-assisted decision tasks. The mathematical ingredients of loss and recovery are not unique to AI; the novelty lies in an AI-specific measurement architecture that turns variation in AI assistance quality into a controlled stressor on human function.

We develop that architecture in two forms. A trajectory form measures degradation, accumulated functional loss, and recovery after a sustained change in AI reliability. A decision-episode form measures performance on cases where AI advice is known to be wrong by comparing the human judgment before advice with the final judgment after advice. Four construct-validation tests draw on five public human--AI studies with 1,074 analytic participants: chess supplies a controlled temporal reliability shift; art and dermatology jointly test the same decision-level definition across nonexpert and professional domains; trip planning separates challenged performance from baseline competence; and logical reasoning tests whether improved calibration necessarily produces improved functional preservation. Table~\ref{tab:validation-logic} states these tests in operational form.

The contribution is fourfold. First, the paper defines demonstrated resilience as a challenge-conditioned property of human--AI task performance rather than a trait, attitude, or generic accuracy measure. Second, it provides an AI-specific measurement architecture at two resolutions: a trajectory form for sustained reliability shifts and a decision-episode form for identifiable misleading-AI events. Third, four validation tests across five studies provide complementary evidence for responsiveness to AI reliability loss, cross-domain interpretability, separation from baseline competence, and distinction from calibration. Fourth, the framework converts the construct into a prospective HCI stress-test protocol for comparing interfaces under the same AI reliability challenge.

The paper therefore connects two previously separated conversations. Research on human resilience and human--autonomy teams establishes why adaptive functioning under technological disruption matters; research on trust and appropriate reliance explains how people use AI advice. The Resilience Stress Test adds the missing functional endpoint: what remains of human task performance when AI quality changes.

\begin{table}[htbp]
\centering
\caption{Construct-validation logic for the Resilience Stress Test.}
\label{tab:validation-logic}
\small
\setlength{\tabcolsep}{5pt}
\renewcommand{\arraystretch}{1.18}
\begin{tabularx}{\textwidth}{p{1.2cm} p{3.7cm} Y}
\toprule
\textbf{Test} & \textbf{Question} & \textbf{Evidence sought} \\
\midrule
V1 & Does the measure respond to an exogenous reliability shift? & Functional degradation should follow the known drop in AI quality, with unaided performance serving as a placebo check. \\
V2 & Does the decision-level definition travel across domains? & Challenged performance should remain interpretable in nonexpert and professional decision tasks. \\
V3 & Can the measure separate resilience from baseline competence? & Apparent condition differences should be re-evaluated after initial challenged performance is accounted for. \\
V4 & Can the measure separate process improvement from functional outcome? & A calibration intervention can be evaluated against challenged task performance in the next batch. \\
\bottomrule
\end{tabularx}
\end{table}

\section{From Reliance to Resilience in Human--AI Interaction}

\subsection{Reliance explains use; resilience evaluates function}

Human factors research has long examined imperfect automation. Misuse and disuse describe problematic patterns of automation use \parencite{parasuraman1997automation}, while trust calibration explains whether reliance corresponds to automation capability \parencite{lee2004trust}. Automation bias research shows how incorrect decision support can propagate into human judgment \parencite{goddard2012automationbias}. Contemporary AI studies make these mechanisms directly observable through an initial decision and a final decision after advice. Relative AI Reliance (RAIR) measures beneficial uptake of correct advice after an initial error, while Relative Self-Reliance (RSR) measures preservation of a correct human judgment when AI advice is wrong \parencite{schemmer2023reliance}. These measures give the decision-level mechanisms that team-oriented resilience research generally does not observe directly.

These measures answer a process question: how did the person use the AI? Resilience answers an outcome question under a defined challenge. When AI advice is wrong, the protected function may be final decision quality. When AI reliability changes over time, the protected function may be sustained task performance. The same reliance tendency can help under one advice state and hurt under another. A user who resists AI strongly may avoid misleading advice and also miss useful corrections. Functional preservation therefore needs its own measurement layer.

Recent work reinforces this point. \textcite{tatasciore2025calibrating} found that greater transparency improved automation-use accuracy, whereas trust-calibration feedback contributed little additional accuracy. \textcite{nair2025solicit} found that solicited AI assistance reduced susceptibility to misleading advice relative to unsolicited assistance. Cognitive-forcing interventions can also reduce overreliance by changing how users engage with AI advice \parencite{bucinca2021trust}. These studies show that interface design can alter advice use. The stress-test perspective asks whether that altered interaction preserves the task function when the AI is wrong.

\subsection{Why unreliable AI creates a distinct HCI stressor}

AI reliability failures differ from ordinary task difficulty in a way that is important for measurement. Task difficulty lowers the quality of the evidence available to the human. Unreliable assistance can instead insert a persuasive but incorrect recommendation into the interaction after the human has already formed an initial judgment. The user must then decide whether to preserve, revise, or abandon that judgment. The resulting error can therefore arise from the task, from the AI, or from the interaction between them. Recording only the final answer obscures these sources.

This distinction becomes more consequential when reliability changes over time. A user may learn a pattern of useful assistance during early interaction and carry that reliance policy into a later period in which the AI becomes less dependable. The failure is then partly temporal: behavior that was adaptive under one reliability regime can become harmful after the regime changes. Human-centered AI research has emphasized the importance of reliable systems, meaningful human control, and methods that account for characteristics unique to AI interaction \parencite{shneiderman2020hcai,xu2022transitioning}. The resilience construct proposed here addresses the complementary question of what the human--AI system achieves after reliability departs from the condition in which that interaction policy was formed.

This is also why resilience should not be reduced to resistance to AI. Refusing advice preserves performance when the AI is wrong but can sacrifice gains when the AI is correct. The art results later in this paper make that trade-off visible: participants who preserved more performance on misleading-AI trials tended to perform worse on the separate correct-AI trials. Resilient interaction therefore requires selective adaptation to changing evidence, not uniformly high or low reliance.

\subsection{Construct boundary and nomological role}

The proposed construct is intentionally narrow. Demonstrated resilience is not a stable personality characteristic, and the stress test does not claim to estimate a person's general capacity to cope with adversity. It is conditional on a protected function, a challenge protocol, and an observation window. A participant can therefore be resilient in one human--AI task and vulnerable in another without contradiction. This conditional definition follows dynamic resilience approaches that infer resilience from functioning in relation to adversity rather than from self-description alone \parencite{kalisch2017framework,kalisch2021freshmo}.

The construct also differs from robustness in interpretation. Preserving performance immediately after a challenge is one resilient response, but recovery after an initial loss is another. A trajectory measure should therefore retain both the depth and duration of disruption. This is the reason for using accumulated functional loss rather than only the minimum post-challenge score. The framework remains compatible with resilience-engineering accounts in which adaptive capacity becomes visible when operating conditions exceed the assumptions under which normal performance was established \parencite{woods2015four,woods2018graceful}.

The expected relationships with neighboring HCI measures follow from these boundaries. Higher baseline competence creates more demonstrated function to preserve, but it does not guarantee preservation after misleading advice. Appropriate reliance can be a mechanism of resilience because accepting correct advice and rejecting incorrect advice protects final performance. Trust and calibration can shape that mechanism by affecting whether the user treats AI output as credible. None of these variables, however, is definitionally equivalent to the challenged outcome. This separation permits an intervention to improve calibration without improving challenged performance, or to change reliance behavior while leaving final function stable. V3 and V4 test precisely these discriminations.

\subsection{Resilience becomes observable under adversity}

Psychological resilience scales estimate perceived coping or recovery capacity \parencite{connor2003cdrisc,smith2008brs}. Dynamic resilience theory shifts attention to observed functioning under stressor exposure \parencite{kalisch2017framework,kalisch2021freshmo}. That shift is especially useful for AI-supported work because reliability can be manipulated experimentally and advice correctness can be identified trial by trial.

Performance-based resilience research provides the mathematical precedent. \textcite{bruneau2003seismic} formalized system response around disruption, and \textcite{poulin2021curves} show how resilience curves capture the depth and duration of functional loss. Human--AI interaction adds direct observations of advice use and decision revision. This makes it possible to measure the outcome trajectory and retain interaction processes as explanatory variables.

Table~\ref{tab:constructs} summarizes the measurement boundary. The stress test complements established HCI constructs by anchoring interpretation to a protected function under a specified challenge.

\begin{table}[htbp]
\centering
\caption{Measurement roles in AI-assisted decision making.}
\label{tab:constructs}
\small
\setlength{\tabcolsep}{5pt}
\renewcommand{\arraystretch}{1.18}
\begin{tabularx}{\textwidth}{p{2.8cm} p{4.3cm} Y}
\toprule
\textbf{Construct} & \textbf{Observed quantity} & \textbf{Role in evaluation} \\
\midrule
Baseline competence & Human task performance before challenged advice & Establishes the function available to be preserved. \\
Trust and calibration & Attitude toward AI and its correspondence with capability & Characterizes the user's representation of system quality. \\
Reliance & Acceptance or rejection of AI advice & Explains how AI input enters the final decision. \\
Demonstrated resilience & Functional performance under a specified AI challenge & Quantifies preservation and recovery of the protected task function. \\
\bottomrule
\end{tabularx}
\end{table}

\section{Resilience Stress Test}

\subsection{Functional trajectory}

A stress test begins by naming the human function whose continuity matters. The challenge protocol \(\chi\) defines the reliability disturbance and the observation window fixes the period over which response is assessed. For human unit \(u\), the variable \(q_u(t)\) records performance of that function. Let \(q_{u,0}>0\) denote performance before challenge onset. Performance is normalized as
\begin{equation}
 p_u(t)=\frac{q_u(t)}{q_{u,0}}.
 \label{eq:normalized}
\end{equation}
This transformation expresses subsequent performance relative to the function demonstrated before the challenge. In repeated binary tasks, \(q_u(t)\) is estimated over a prespecified block or model-based probability so that the trajectory reflects a performance level.

The deepest observed degradation is
\begin{equation}
D_u=\left[1-\min_t p_u(t)\right]_+,
\label{eq:degradation}
\end{equation}
where \([x]_+=\max(x,0)\). The average accumulated deficit over an observation interval of length \(T\) is
\begin{equation}
L_u=\frac{1}{T}\int [1-p_u(t)]_+\,dt,
\qquad
FR_u=1-L_u.
\label{eq:fr}
\end{equation}
The \(FR\) transformation places stronger preservation at larger values. It inherits the area-of-loss logic used in resilience-curve analysis and adapts it to normalized human performance \parencite{bruneau2003seismic,poulin2021curves}. This choice gives duration the same conceptual status as depth: a brief performance dip and a persistent collapse can share the same minimum while producing different accumulated loss. For equally spaced discrete windows,
\begin{equation}
\widehat{FR}_u
=1-\frac{1}{K}\sum_{k=1}^{K}[1-p_{u,k}]_+ .
\label{eq:discrete-fr}
\end{equation}
Recovery can then be reported as the earliest sustained return to a prespecified fraction of baseline. The chess analysis uses 90\% as a descriptive recovery threshold because the post-change observation period contains only ten trials.

The baseline requires adequate performance to support ratio normalization. A very small denominator can create unstable resilience values. The chess analysis therefore uses \(q_{\min}=.30\) for participant-level normalized summaries and reports the raw trajectory for all participants. The threshold is an operational safeguard for this dataset. Sensitivity to baseline definition is examined explicitly because resilience magnitude depends on the reference function being protected.

\subsection{Decision-level challenge episodes}

Several public human--AI datasets contain sparse misleading-advice episodes in place of a sustained temporal shock. For participant \(i\) and decision \(j\), let \(b_{ij}\) indicate whether the initial human decision is correct, \(a_{ij}\) whether the AI advice is correct, and \(f_{ij}\) whether the final decision is correct. Misleading-AI trials satisfy \(a_{ij}=0\). Within those trials,
\begin{equation}
B_i^- = \mathbb{E}(b_{ij}\mid a_{ij}=0),
\qquad
CP_i^- = \mathbb{E}(f_{ij}\mid a_{ij}=0),
\label{eq:cp}
\end{equation}
and
\begin{equation}
\Delta_i^- = CP_i^- - B_i^- .
\label{eq:delta}
\end{equation}
\(B^-\) is initial competence on the challenged cases. \(CP^-\) is the protected functional outcome after exposure to misleading AI. \(\Delta^-\) measures the change associated with the advice episode. The three quantities answer different measurement questions. \(B^-\) locates the starting function on the cases where AI fails. \(CP^-\) shows what survives after advice. \(\Delta^-\) isolates the within-episode change, which makes task difficulty visible in the resilience interpretation.

RAIR and RSR remain useful process measures. RAIR captures correction of an initial human error when AI advice is correct. RSR captures preservation of an initially correct human judgment when AI advice is wrong. Participant ratios are reported only when the eligible denominator is positive. This denominator rule keeps the ratio tied to observable opportunities for reliance.

\subsection{What is AI-specific about the metrics}

The underlying resilience arithmetic is deliberately generic: normalized loss and recovery can be used in other domains. What makes the present metrics specific to human--AI interaction is how the stressor, observations, and interpretation are defined. The challenge is a change in AI assistance quality---for example, a reliability drop or an identifiable wrong recommendation---while the protected outcome remains human task performance. This separates loss caused by the AI-mediated interaction from ordinary variation in task difficulty.

The decision-level formulation uses a structure that is especially characteristic of AI advice systems: an initial human judgment, an AI recommendation with known correctness, and a final human judgment. Conditioning on misleading-AI trials makes \(B^-\) the human's competence on exactly the cases where the AI fails, \(CP^-\) the function that survives after exposure to that failure, and \(\Delta^-\) the within-case change following the advice episode. The same episode can also be paired with RAIR and RSR to explain whether preservation arose from accepting or rejecting AI input.

AI assistance also creates a distinctive two-sided evaluation problem. Uniform resistance to AI can look resilient when advice is wrong but can destroy the benefit of correct advice. A valid human--AI resilience assessment must therefore report challenged preservation together with normal-condition benefit. The metrics are AI-specific in this operational sense: they quantify how human function changes because AI assistance changes quality, rather than treating resilience as a general personal capacity.

\subsection{Interpretation rules for a stress test}

A resilience score is meaningful only relative to the challenge that generated it. Reporting \(FR=.8\) without the baseline, reliability shift, task, and observation window would invite false comparisons. A complete stress-test result should therefore identify five elements: the protected function \(q\), the baseline used to estimate \(q_0\), the challenge protocol \(\chi\), the post-challenge observation window, and the process measures collected to explain the response. These elements define the test condition in the same way that workload, display, and task conditions define a conventional HCI evaluation.

The trajectory and decision-level forms serve different experimental structures. The trajectory form is appropriate when reliability changes for long enough to observe degradation and adaptation. It supports \(D\), \(FR\), and a recovery measure. The decision-level form is appropriate when incorrect AI outputs are embedded as identifiable episodes among otherwise useful recommendations. It supports \(B^-\), \(CP^-\), and \(\Delta^-\), with RAIR and RSR retained as mechanism measures. The two forms should not be collapsed into one universal scalar because they preserve different information about the interaction.

For between-interface comparisons, the challenge should be held constant. Suppose two explanation designs yield the same nominal accuracy when the AI is reliable. If one design produces a smaller post-change loss or faster recovery after the same reliability decline, the designs are equivalent under nominal evaluation but different under resilience evaluation. Conversely, a design that increases resistance to bad advice at the cost of systematically ignoring correct advice may improve \(CP^-\) while lowering performance in normal operation. Reporting both nominal benefit and challenged preservation prevents a resilience intervention from being rewarded for simply suppressing AI use.

\begin{table}[htbp]
\centering
\caption{Recommended elements of a prospective Human--AI Resilience Stress Test.}
\label{tab:protocol}
\small
\setlength{\tabcolsep}{4pt}
\renewcommand{\arraystretch}{1.18}
\begin{tabularx}{\textwidth}{p{2.5cm} p{4.5cm} Y}
\toprule
\textbf{Element} & \textbf{Operational choice} & \textbf{Reason for reporting} \\
\midrule
Protected function & Accuracy, decision quality, throughput, or another task outcome with direct human value & States what the interaction is expected to preserve. \\
Baseline & Unaided or stable prechallenge performance measured on comparable cases & Separates available competence from challenge-induced loss. \\
AI challenge & Prespecified change in correctness, uncertainty, or error regime & Makes the stressor reproducible and comparable across interfaces. \\
Trajectory window & Trial blocks or time windows long enough to observe early loss and later adaptation & Distinguishes transient disruption from persistent loss. \\
Process measures & Trust, confidence, advice acceptance, RAIR/RSR, response time, or explanation use & Explains how the observed functional response arose. \\
Normal-condition outcome & Performance on reliable or correct-AI trials & Detects interventions that gain resilience by discarding useful AI assistance. \\
\bottomrule
\end{tabularx}
\end{table}

\section{Multi-Study Validation Across Five Human--AI Studies}

\subsection{Datasets and source reconstruction}

A targeted search of public human--AI decision datasets and associated publications was conducted through August 2026. Eligible datasets contained ground-truth task performance and observable human decisions around AI advice. The final set contains five studies and 1,074 analytic participants (Table~\ref{tab:datasets}). These five source studies support four validation tests because the art and dermatology studies jointly form V2, the cross-domain test. A candidate dataset with fine-grained fact-checking judgments was excluded from empirical analysis because its repository was under embargo when the analysis package was frozen.

\begin{table}[htbp]
\centering
\caption{Public datasets used for multi-study construct validation.}
\label{tab:datasets}
\small
\setlength{\tabcolsep}{4pt}
\renewcommand{\arraystretch}{1.18}
\begin{tabularx}{\textwidth}{p{2.6cm} p{0.9cm} p{4.0cm} Y}
\toprule
\textbf{Study} & \textbf{N} & \textbf{Interaction structure} & \textbf{Validation role} \\
\midrule
\textcite{chong2022confidence,chong2023data} & 100 & Thirty chess trials; AI quality changes after trial 20 & Dynamic response to a known reliability shift \\
\textcite{kuper2024traits} & 244 & Twenty-four art judgments with five misleading-AI trials & Cross-domain challenged performance \\
\textcite{kuper2025derma} & 223 & Twenty-four dermatology judgments with five misleading-AI trials & Professional-domain challenged performance \\
\textcite{salimzadeh2024uncertainty} & 258 & Trip planning in a 3\(\times\)2 complexity--uncertainty design & Baseline competence and challenge sensitivity \\
\textcite{he2023knowing} & 249 & Logical reasoning across two task batches with tutorial and XAI conditions & Intervention responsiveness \\
\bottomrule
\end{tabularx}
\end{table}

Source reconstruction preceded each new analysis. In chess, the released manipulation and the published acceptance model were reproduced. The art and dermatology files matched the released accuracy and reliance variables exactly. Trip-planning reconstruction recovered the 258 participants, the six equally sized experimental cells, and the intended pattern of correct versus misleading AI advice. The logical-reasoning files reproduced the published tutorial analyses, including the calibration result. These checks establish that the new resilience variables are computed from the same decisions that support the source findings.

The source populations differ substantially. The art study retained 244 participants after attention-check screening and had a mean age of 29.91 years \parencite{kuper2024traits}. The dermatology analysis retained 223 participants with a mean age of 44.3 years and mean dermatology experience of 15.5 years \parencite{kuper2025derma}. Trip planning retained 258 Prolific participants \parencite{salimzadeh2024uncertainty}, and the logical-reasoning study analyzed 249 Prolific participants \parencite{he2023knowing}. Each dataset is analyzed within study because the tasks and participant populations define different interaction contexts.

\subsection{Analysis strategy}

V1 uses the chess reliability shift as the strongest dynamic test. Final advantageous-move correctness is modeled at the trial level with trial fixed effects and participant-clustered standard errors. The key interaction compares the post-change response of the 80\%\(\rightarrow\)20\% condition with the 20\%\(\rightarrow\)80\% directional control. Unaided initial correctness provides a placebo outcome. For the degradation condition, trials 11--20 define the primary baseline and trials 21--25 and 26--30 define the two post-change windows. Trials 1--20 provide a baseline sensitivity analysis.

V2 applies identical \(B^-\), \(CP^-\), and \(\Delta^-\) definitions to art and dermatology. Participant bootstrap intervals quantify uncertainty. RAIR and RSR describe reliance mechanisms using positive eligible denominators. Associations with performance on the 19 correct-AI trials use a criterion that excludes the five misleading-AI cases used to define challenged performance.

V3 estimates the six trip-planning cells created by three complexity levels and two uncertainty types. Robust participant-level models evaluate \(CP^-\) before and after adjustment for \(B^-\), then evaluate \(\Delta^-\). A trial-level linear-probability robustness model uses participant-clustered standard errors because the fully interacted logistic specification exhibits separation in these data.

V4 first reproduces the source tutorial result. The new analysis models final correctness on misleading-AI trials across batches while controlling initial correctness and task identity. Participant-clustered standard errors account for repeated decisions. The Batch~2\(\times\)Tutorial term estimates intervention responsiveness, and the Batch~2\(\times\)XAI term is included in the same model. Calibration-group heterogeneity is evaluated on participant change scores.

The analyses use public datasets for construct validation across known manipulations and neighboring constructs. Because the datasets were inspected during method development, interpretation centers on effect sizes, confidence intervals, source reconstruction, and the consistency of the pattern across designs rather than on any single significance test. V1 supplies the strongest exogenous reliability test, while V2--V4 test portability and separation from neighboring explanations.

The validation logic emphasizes four forms of evidence rather than a single significance test. \emph{Responsiveness} requires challenged performance to change when AI reliability changes exogenously. \emph{Cross-context interpretability} requires the same operational definition to retain a clear meaning across tasks. \emph{Discriminant evidence} requires the resilience outcome to remain distinguishable from baseline competence and process variables such as calibration. \emph{Mechanistic coherence} requires the observed pattern to be interpretable using trial-level interaction behavior rather than only a post hoc label. No one public dataset contains all four features, which motivates the multi-study design.

\section{Results}

\subsection{V1: A reliability drop produced immediate functional loss and partial adaptation}

\begin{figure}[htbp]
\centering
\includegraphics[width=\textwidth]{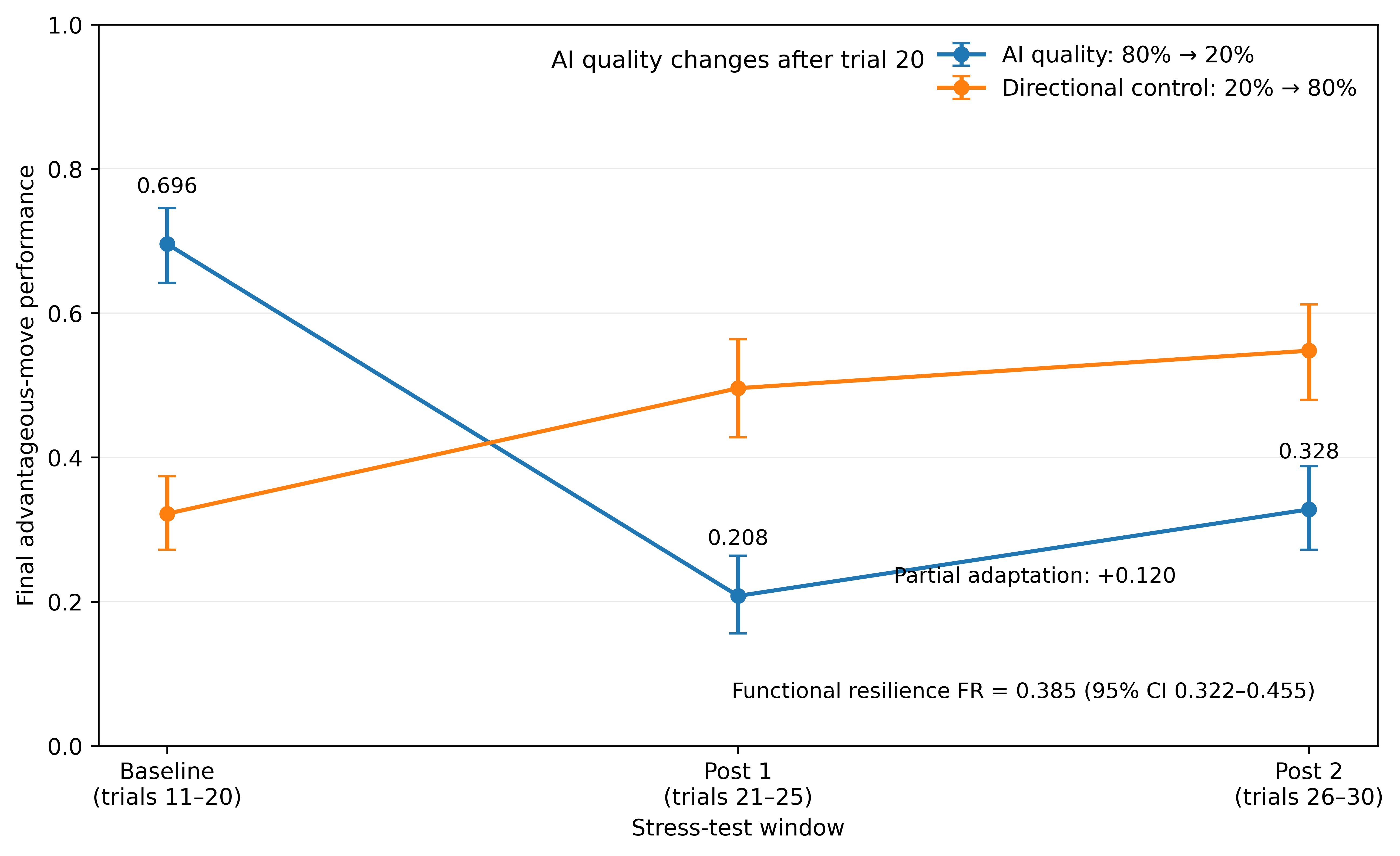}
\caption{Dynamic stress-test response to a controlled AI-quality shift. Points show final advantageous-move performance across the immediate pre-change window and two post-change windows; error bars are participant-bootstrap 95\% confidence intervals. The degradation condition shows a sharp loss followed by partial adaptation.}
\label{fig:v1}
\end{figure}

The known deterioration in AI quality generated the trajectory expected from a valid stress-test measure. In the degradation group, final advantageous-move performance was .696 during trials 11--20, fell to .208 in trials 21--25, and reached .328 in trials 26--30 (Figure~\ref{fig:v1}). Equation~\ref{eq:discrete-fr} yielded \(FR=.385\), bootstrap 95\% CI [.322, .455], with maximum degradation \(D=.701\). Performance improved by .120 from the first post-change window to the second, 95\% CI [.052, .188], indicating partial adaptation during continued exposure to low-quality AI.

The direction-by-post interaction was large (OR=.096, 95\% CI [.064, .144], \(p<.001\)), and the participant-fixed-effects robustness model produced OR=.077 (\(p<.001\)). The corresponding interaction for unaided initial correctness was small (\(p=.670\)). This placebo result localizes the disruption to the AI-mediated decision stage. Among the 48 degradation-group participants meeting \(q_{\min}=.30\), mean participant \(FR=.381\), and 8.3\% reached at least 90\% of their own baseline in the final window. Using trials 1--20 as baseline yielded \(FR=.443\), 95\% CI [.372, .518], confirming the same degradation pattern with a less demanding reference level.

\subsection{V2: Misleading AI reduced performance in both domains, with different reliance profiles}

The same decision-level definition captured functional loss in a nonexpert art task and a professional dermatology task. On the five misleading-AI art trials, \(B^-=.512\), \(CP^-=.340\), and \(\Delta^-=-.172\), 95\% CI [-.202, -.143]. Dermatology showed \(B^-=.510\), \(CP^-=.439\), and \(\Delta^-=-.071\), 95\% CI [-.090, -.053]. The smaller decrease in dermatology is a descriptive cross-study difference because the task and participant populations differ.

Reliance behavior varied more sharply than challenged performance. Mean participant RAIR was .306 in art and .100 in dermatology. Mean participant RSR was .666 in art and .851 in dermatology. In art, higher \(CP^-\) was associated with lower final accuracy on the 19 correct-AI trials (\(\rho=-.419\), 95\% CI [-.518, -.309], \(p<.001\)); \(\Delta^-\) showed an even stronger negative association (\(\rho=-.613\), 95\% CI [-.688, -.527], \(p<.001\)). Dermatology showed weaker associations. The pattern reveals selective reliance as the relevant interaction problem. Preserving judgment under misleading advice requires resistance to a bad recommendation. Benefiting from correct advice requires receptivity when the recommendation adds value.

\subsection{V3: Baseline adjustment substantially attenuated the apparent complexity gradient}

\begin{figure}[htbp]
\centering
\includegraphics[width=0.72\textwidth]{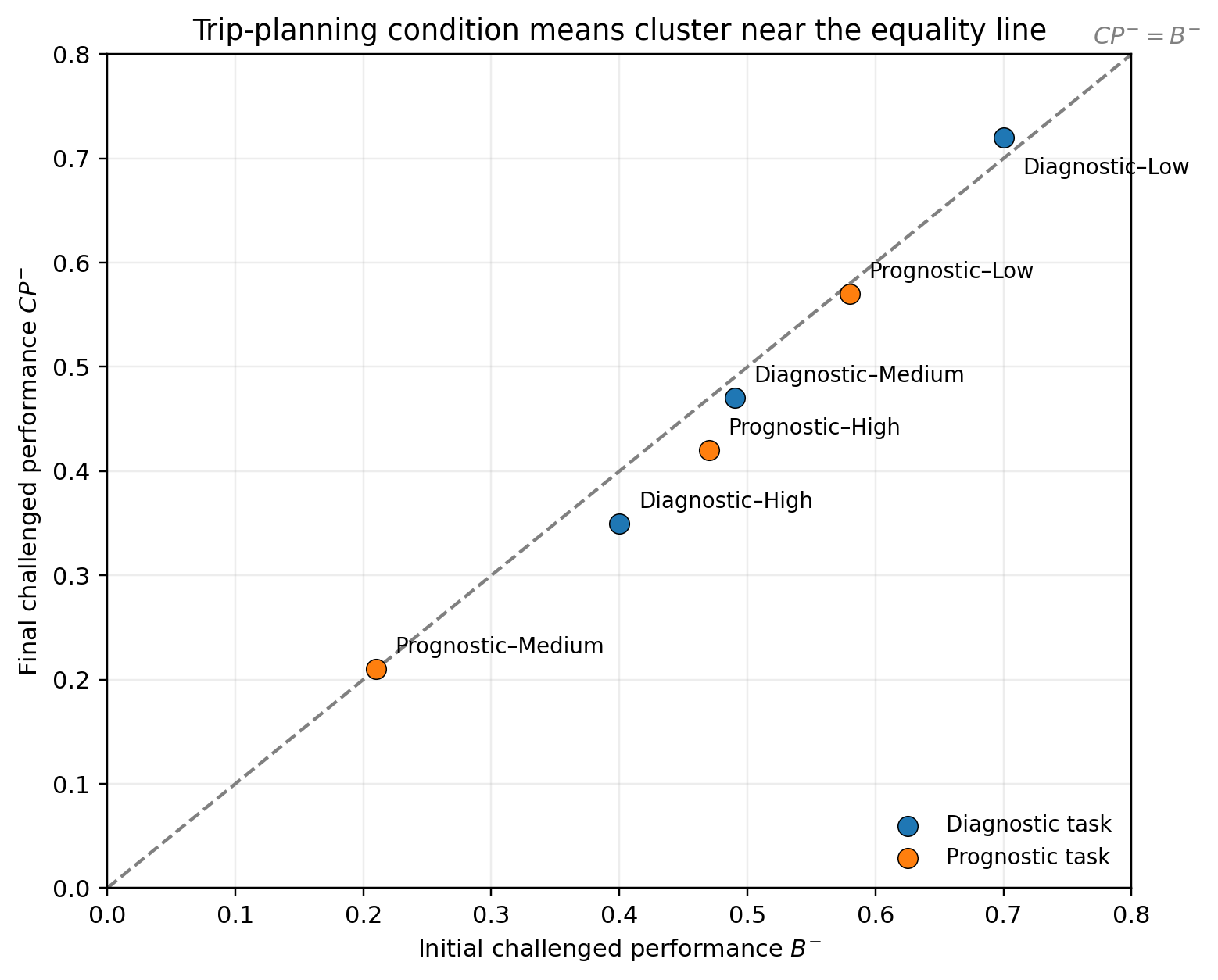}
\caption{Relationship between initial challenged performance ($B^{-}$) and final challenged performance ($CP^{-}$) across the six trip-planning conditions. Each point is a condition mean labeled by task type (diagnostic or prognostic) and complexity level. The dashed 45$^{\circ}$ line marks equality ($CP^{-}=B^{-}$). The close alignment of the six cell means with this line shows that much of the raw variation in final challenged performance was already present before exposure to misleading AI, which is why the apparent complexity effect attenuates after adjustment for $B^{-}$.}
\label{fig:v3}
\end{figure}

Trip-planning challenged performance varied strongly across the six experimental cells before baseline adjustment. Raw \(CP^-\) ranged from .209 to .721, and complexity was associated with unadjusted \(CP^-\) (Wald \(\chi^2=14.02\), \(p<.001\)). Figure~\ref{fig:v3} shows that the six cell means lie close to the \(CP^- = B^-\) reference line, indicating that final challenged performance closely tracked baseline challenged performance. After adjustment for \(B^-\), the complexity statistic fell to \(\chi^2=4.26\) with \(p=.119\). Across the equally sized cells, mean \(B^-=.473\), \(CP^-=.457\), and \(\Delta^-=-.016\). The \(\Delta^-\) models produced weak evidence for complexity (\(p=.170\)), uncertainty (\(p=1.000\)), or their interaction (\(p=.377\)).

This result has a direct HCI measurement implication. Final performance under difficult AI-assisted conditions can look like vulnerability when substantial variation was already present in the human's initial ability on those same cases. Capturing the initial decision provides the reference needed to identify the change associated with misleading AI.

\subsection{V4: Calibration improved while challenged performance remained stable}

The source tutorial produced the expected change in self-assessment. Among tutorial participants who were initially miscalibrated, absolute self-assessment error decreased from 1.67 to 1.14 (\(p<.001\)), reproducing the original result \parencite{he2023knowing}. The challenged functional outcome showed a different pattern. The Batch~2\(\times\)Tutorial estimate on misleading-AI final correctness was \(\beta=-.014\), 95\% CI [-.089, .060], \(p=.705\). The Batch~2\(\times\)XAI estimate was near zero (\(p=.981\)), and the task-adjusted batch estimate for unaided initial correctness was also small (\(p=.283\)). Calibration-group heterogeneity in \(CP^-\) and \(\Delta^-\) change was weak.

The result separates an interaction process from the protected outcome. The tutorial sharpened participants' assessment of their own performance, while preservation under misleading AI changed little on average. For HCI evaluation, calibration therefore provides diagnostic information about the user's internal model of performance; challenged task performance provides evidence about whether the function survives erroneous AI assistance.

\begin{table}[htbp]
\centering
\caption{Summary of construct-validation results.}
\label{tab:results}
\small
\setlength{\tabcolsep}{4pt}
\renewcommand{\arraystretch}{1.18}
\begin{tabularx}{\textwidth}{p{2.2cm} p{5.1cm} Y}
\toprule
\textbf{Validation} & \textbf{Principal result} & \textbf{Measurement implication} \\
\midrule
V1 Chess & \(FR=.385\), 95\% CI [.322, .455]; direction-by-post OR=.096 & The measure responds strongly to an exogenous reliability decline and captures partial adaptation. \\
V2 Art / dermatology & \(\Delta^-=-.172\) and -.071, with domain-specific reliance profiles & Challenged performance travels across distinct decision domains and remains separable from reliance. \\
V3 Trip planning & Raw complexity \(p<.001\); baseline-adjusted complexity \(p=.119\) & Initial task competence is required for interpreting final challenged performance. \\
V4 Logical reasoning & Calibration improved; Batch~2\(\times\)Tutorial \(\beta=-.014\), \(p=.705\) & Process improvement and functional preservation can move independently. \\
\bottomrule
\end{tabularx}
\end{table}

\section{Discussion: What Resilience Adds to Human--AI Evaluation}

\subsection{A functional endpoint under AI failure}

The main contribution is methodological. Nominal accuracy describes performance when AI assistance behaves as expected; the Resilience Stress Test adds a controlled reliability challenge and measures how much human task function is preserved, lost, and recovered. This creates a functional endpoint for failure conditions while trust, calibration, RAIR, and RSR remain explanatory process measures.

The four validation tests, drawing on five studies, show why this additional layer matters. In chess, AI-mediated performance collapsed after the reliability shift even though unaided task ability did not show the same disruption. In trip planning, adjustment for baseline competence substantially attenuated the apparent complexity gradient in challenged performance. Art and logical reasoning further show that resistance to misleading advice, benefit from correct advice, and calibration need not move together. A human--AI evaluation can therefore distinguish the function that survives from the interaction mechanism that produced that outcome.

This framing also clarifies the relationship to appropriate reliance. Appropriate reliance can support resilience, but it is a mechanism rather than the protected outcome itself. The stress-test endpoint remains anchored to what the human--AI system is intended to accomplish when AI quality changes.

\subsection{Designing and comparing interfaces under reliability stress}

The framework changes the interface question from whether a design improves average trust or accuracy to whether it helps preserve task function when AI quality changes, and at what cost under normal operation. Competing interfaces can face the same prespecified reliability shift and be compared on accumulated functional loss, challenged performance, and recovery, while confidence, advice acceptance, response time, or explanation use identify the mechanism.

This comparison can reveal trade-offs hidden by nominal testing. An explanation may slightly slow normal decisions yet reduce loss after a reliability decline. A forcing function may protect against wrong advice while also suppressing useful AI corrections. A confidence display may improve calibration without improving challenged performance. Reporting both normal-condition benefit and challenged preservation prevents an intervention from appearing resilient merely because it discourages AI use.

The stress test is therefore directly usable as an HCI experimental protocol. A study can establish a stable prechallenge function, introduce a deployment-relevant reliability disturbance while holding the task as constant as possible, and observe both early disruption and later adaptation. The challenge can be tailored to the application---diagnostic errors in clinical support, unsupported factual suggestions in writing assistance, or recommendation-quality shifts in planning systems---while preserving a common evaluation logic.

\subsection{What the four validation tests establish}

The four validation tests draw on five studies and provide complementary forms of construct evidence. Chess establishes responsiveness to an exogenous reliability shift and demonstrates temporal degradation and partial adaptation. Art and dermatology show that the same decision-level definition remains interpretable across nonexpert and professional domains. Trip planning demonstrates why case-matched baseline competence is necessary to distinguish low starting ability from challenge-induced loss. Logical reasoning shows that a neighboring process construct can improve while challenged task performance remains stable. Together, these results support demonstrated resilience as a distinct, challenge-conditioned property of human--AI performance.

The evidence also defines a clear next step. The present datasets establish the measurement logic across several existing experimental structures; prospective studies can now prespecify the protected function, AI challenge, observation window, and primary resilience outcome. Such studies can estimate participant-level reliability with more challenged episodes and test interface interventions directly against the same stress protocol. The contribution of the present paper is to provide the construct, metrics, validation logic, and experimental protocol needed to make those comparisons possible.

\section{Extension Beyond the Individual Decision}

AI-supported work often connects individual decisions to collective outcomes. The same stress-test logic can be applied at the level where the protected function exists. For a team, $q(t)$ may represent coordinated output, decision quality, or task completion under a shared AI failure. For an organization, it may represent continuity of an AI-supported service or workflow. Individual resilience can then be modeled as one contributor to the higher-level trajectory rather than being mechanically averaged into a collective score, consistent with multilevel resilience theory \parencite{fisher2023multilevel}.

This extension is especially relevant because AI can create common-mode failures. A shared model may provide the same misleading recommendation to many team members, an interface may direct attention toward the same incorrect cue, or a workflow may remove independent checks after a period of strong AI performance. A team or organizational stress test can measure whether coordination, redundancy, escalation, and human override preserve the collective function under that shared challenge. The framework therefore provides a common measurement logic from individual decision support to broader human--AI work systems.

\section{Conclusion}

This paper contributes a new measurement framework for evaluating human performance when AI assistance becomes unreliable. First, it defines \emph{demonstrated resilience} as preservation or recovery of human task function under a specified AI reliability challenge, distinguishing the construct from baseline competence, trust, calibration, and reliance. Second, it operationalizes resilience at two complementary resolutions: trajectory-based functional resilience for sustained reliability shifts and decision-level $B^-$, $CP^-$, and $\Delta^-$ for identifiable misleading-AI episodes. Their AI-specificity comes from conditioning human performance on AI assistance quality and, at the decision level, exploiting the sequence of pre-AI judgment, known AI advice correctness, and post-AI judgment. Third, four validation tests drawing on five public studies with 1,074 participants provide complementary evidence for responsiveness, cross-domain interpretability, separation from baseline competence, and distinction from calibration. Fourth, the paper converts these insights into a prospective HCI stress-test protocol for comparing interfaces under reproducible AI failure conditions.

The broader contribution is a shift in what human--AI evaluation treats as success. Reliable performance when AI works as expected remains important, but systems intended for real-world use should also be judged by how much human function survives when AI quality changes. Measuring normal-condition benefit together with challenged preservation makes that property observable. In this form, resilience becomes both a measurable characteristic of human--AI interaction and a concrete design target for AI systems expected to remain useful under imperfect assistance.

\section*{Disclosure statement}
The author declares no competing interests.

\section*{Data availability statement}

This study analyzes publicly available human--AI interaction datasets. The chess data are available from Mendeley Data at \url{https://doi.org/10.17632/ng33vg479n.1}. The art and dermatology data are available from the Open Science Framework project 87tpu at \url{https://osf.io/87tpu/}, in the \texttt{Dataset\_Art} and \texttt{Dataset\_Derma} subfolders. The trip-planning data are available from 4TU.ResearchData at \url{https://doi.org/10.4121/76D36276-9D77-4CAD-95B5-C866EF196F89}. The logical-reasoning data are available from 4TU.ResearchData at \url{https://doi.org/10.4121/96010177-46E8-4967-9A49-FE38F0BACE4E}. No new participant data were collected for this study.

\section*{Ethics statement}
This study reanalyzed de-identified public datasets and conducted no new participant recruitment or data collection. Ethical approval and informed-consent procedures for the original studies are reported in their respective source publications.

\section*{Declaration of generative AI and AI-assisted technologies in the writing process}
OpenAI's ChatGPT was used to support language editing, manuscript organization, and background research. 

\section*{Notes on contributor}
\textbf{Shaoshan Liu} is Director of Embodied AI at the Shenzhen Institute of Artificial Intelligence and Robotics for Society (AIRS). His research spans human--AI systems, embodied AI, and technology policy. He serves on the ACM U.S. Technology Policy Council and is a member of the Technology Leadership Panel Advisory Group of the U.S. National Academy of Public Administration (NAPA). He received his Ph.D. in Computer Engineering from the University of California, Irvine, and his Master of Public Administration (MPA) from Harvard Kennedy School.

\printbibliography

\end{document}